\documentclass[preprint,aps,prd,superscriptaddress,preprintnumbers,nofootinbib,amsmath,amssymb]{revtex4}
\usepackage{graphicx}

\newcommand\bea{\begin{eqnarray}}
\newcommand\eea{\end{eqnarray}}
\newcommand\ft[2]{{\textstyle\frac{#1}{#2}}}
\newcommand\fft[2]{\frac{#1}{#2}}
\newcommand\nn{{\nonumber}}

\begin{document}
\preprint{MCTP-12-12}

\title{Holographic Lifshitz flows and the null energy condition}

\author{James T.~Liu}
\email{jimliu@umich.edu}
\affiliation{Michigan Center for Theoretical Physics,
Randall Laboratory of Physics,
The University of Michigan,
Ann Arbor, MI 48109--1040, USA}

\author{Zhichen Zhao}
\email{zhichen@umich.edu}
\affiliation{Michigan Center for Theoretical Physics,
Randall Laboratory of Physics,
The University of Michigan,
Ann Arbor, MI 48109--1040, USA}

\begin{abstract}
We study asymptotically Lifshitz spacetimes and the constraints on flows between Lifshitz fixed
points imposed by the null energy condition.  In contrast with the relativistic holographic
$c$-theorem, where the effective AdS radius, $L$, is monotonically decreasing in the IR,
for Lifshitz backgrounds we
find that both $L$ and $z$ may flow in either direction.  We demonstrate this with several
numerical examples in a phenomenological model with a massive gauge field coupled
to a real scalar.
\end{abstract}

\maketitle

\section{Introduction}

It is well known that energy conditions in general relativity play a crucial role in understanding
the global structure of spacetime.  In particular, such conditions are used as inputs to the
Penrose-Hawking singularity theorems as well as the Hawking area theorem of black hole
thermodynamics.  Of particular interest is the null energy condition, which states that
\begin{equation}
T_{\mu\nu}k^\mu k^\nu\ge0,
\end{equation}
for any null vector field $k^\mu$.  While this is a requirement imposed on the matter content
of the theory, application of the Einstein equation
\begin{equation}
R_{\mu\nu}-\ft12g_{\mu\nu} R=8\pi GT_{\mu\nu},
\end{equation}
immediately converts this to a condition on the geometry
\begin{equation}
R_{\mu\nu}k^\mu k^\nu\ge0.
\label{eq:riccicond}
\end{equation}

As an example of how the null energy condition is applied, consider an affinely
parametrized null congruence specified by $k^\mu\partial_\mu=d/d\lambda$.  The
Raychaudhuri equation then gives
\begin{equation}
\fft{d\theta}{d\lambda}=-\fft1{D-2}\theta^2-\sigma_{\mu\nu}\sigma^{\mu\nu}
+\omega_{\mu\nu}\omega^{\mu\nu}-R_{\mu\nu}k^\mu k^\nu,
\end{equation}
where $\theta$ is the expansion of the null congruence, $\sigma_{\mu\nu}$ the shear and
$\omega_{\mu\nu}$ the twist.  (Here $D$ is the dimension of spacetime.)  So long as the
congruence is twist-free and the null energy condition is satisfied, the right hand side is then
non-positive, and we may conclude that $d\theta/d\lambda\le0$.  Moreover, in this case the
inequality $d\theta/d\lambda\le -\theta^2/(D-2)$ may be integrated to demonstrate that any
negative expansion necessarily leads to the formation of caustics.

Turning to AdS/CFT, the null energy condition plays a key role in the proof of the holographic
$c$-theorem concerning flows between the UV and IR
\cite{Alvarez:1998wr,Girardello:1998pd,Freedman:1999gp,Sahakian:1999bd}.  In particular,
working in the Poincar\'e patch and assuming $d$-dimensional Lorentz invariance, the bulk
metric may be parametrized as
\begin{equation}
ds_{d+1}^2=e^{2A}(-dt^2+d\vec x_{d-1}^{\,2})+dr^2,
\label{eq:grel}
\end{equation}
where $A$ is a function of the bulk radial coordinate $r$.  For pure AdS, this function takes
the form $A=r/L_0$, where $L_0$ is the AdS radius.  However, more generally, we may
define an effective AdS radius $L(r)$ along flows according to $A'(r)=1/L(r)$, or equivalently
$L(r)=1/A'(r)$, where primes denote derivatives with respect to $r$.  This effective radius
agrees with the true AdS radius at fixed points of the flow.

The bulk metric (\ref{eq:grel}) has the form of a domain wall solution, and the Ricci tensor is
easily computed:
\begin{equation}
R_{\mu\nu}=-(A''+dA'^2)g_{\mu\nu},\qquad R_{rr}=-d(A''+A'^2).
\end{equation}
Choosing a null vector field $k^\mu\partial_\mu=e^{-A}\partial_t+\partial_r$, the null
energy condition in the bulk then translates to the Ricci condition (\ref{eq:riccicond}), which
reads
\begin{equation}
e^{-2A}R_{tt}+R_{rr}=-(d-1)A''\ge0,
\end{equation}
or equivalently $A''\le0$.  The holographic $c$-theorem immediately follows by taking $A'=1/L$,
so that we are left with the inequality $L'\ge0$.  This demonstrates that the effective AdS radius
is monotonic increasing towards the UV, and furthermore allows us to define a corresponding
monotonic $c$-function
\cite{Alvarez:1998wr,Girardello:1998pd,Freedman:1999gp,Sahakian:1999bd}.

The general idea behind the holographic $c$-theorem is that the Weyl anomaly of the boundary
field theory may be computed in the bulk dual through gravitational methods
\cite{Henningson:1998gx,Henningson:1998ey}.  Since unitarity of the field theory is a
crucial input to both the
two-dimensional Zamoldchikov $c$-theorem \cite{Zamolodchikov:1986gt} and the recently
constructed four-dimensional $a$-theorem \cite{Komargodski:2011vj}, it is natural to expect
a corresponding requirement on the matter content of the bulk theory.  Such a requirement
is naturally aligned with the null energy condition in the bulk, as violations of the null energy
condition will lead to superluminal propagation and instabilities in the bulk
\cite{Gao:2000ga,Buniy:2005vh,Dubovsky:2005xd,Buniy:2006xf}, with corresponding 
violations in the holographic dual \cite{Kleban:2001nh}.

Questions about the nature of the bulk matter content become more pronounced when extending
the holographic $c$-theorem to include higher derivative terms in the bulk theory
\cite{Sinha:2010ai,Oliva:2010eb,Myers:2010xs,Sinha:2010pm,Myers:2010tj,Liu:2010xc,Paulos:2011zu,Liu:2011ii}.
In particular, the null energy condition is no longer directly connected to the Ricci
tensor according to (\ref{eq:riccicond}), as the leading order Einstein equation will now pick up
higher curvature corrections.  Physically, higher derivative gravitational interactions may lead
to non-unitary propagation of bulk gravitational modes, in which case one would hardly expect
the dual field theory to be well behaved.  Thus additional constraints on the higher derivative
terms in the gravity sector must be imposed along with the null energy condition in the matter
sector in order to obtain a well-behaved holographic $c$-function.

While higher curvature bulk theories are a subject of much recent investigations, here we are
instead interested in applying the null energy condition to bulk duals of non-relativistic systems,
and will restrict our focus to Einstein gravity in the bulk.  In particular, holographic techniques
have been developed for the study of strongly coupled critical points in condensed matter
theories (see {\it e.g.}~\cite{Hartnoll:2009sz,Herzog:2009xv,McGreevy:2009xe} for recent reviews). 
In a non-relativistic context, the time and space components of quantum critical systems do
not necessarily exhibit the same scaling symmetry, and Lorentz invariance is broken into
\bea
t \rightarrow \lambda ^z t, \quad \vec x\rightarrow \lambda \vec x,
\eea
under scaling by $\lambda$.   Here $z$ is the dynamical exponent, and Lorentz symmetry
is broken when $z\ne1$.  Asymptotically, this Lifshitz scaling may be realized in the
$(d+1)$-dimensional holographic dual by taking a bulk metric of the form
\bea
ds_{d+1}^2=-e^{2zr/L}dt^2+e^{2r/L}d\vec x_{d-1}^{\,2}+dr^2,
\label{metric}
\eea
where $L$ is the analog of the AdS radius, and parametrizes the bulk curvature.  The Lifshitz
scaling is then realized by shifts in the radial coordinate $r$.  Note that $d$-dimensional
Lorentz invariance is restored when $z=1$, in which case this metric reduces to that of the
Poincar\'e patch of pure AdS.

In this paper, we will explore the implications of the null energy condition on flows between
Lifshitz fixed points as well as flows between AdS and Lifshitz fixed points.  In particular, we
generalize the relativistic $c$-theorem inequality $L'\ge0$ to the Lifshitz case.  Since time
and space scaling are distinct, one may obtain two monotonic $c$-functions constraining the
flows of $L$ and $z$.  However, the interpretation of these monotonic functions is not at all
obvious; it is not clear whether they count degrees of freedom, and indeed their relation to
Lifshitz scaling is obscure.  Moreover, unlike in the relativistic case, we see
that flows towards increasing and decreasing $L$ are both allowed, so long as the dynamical
exponent $z$ is also allowed to flow \cite{Braviner:2011kz}.

In order to get a better understanding of the null energy condition and its relation to Lifshitz
flows, we examine a phenomenological model where the Lifshitz geometry arises from a
massive vector field coupled to a real scalar.  By adjusting the scalar potential as well as the
scalar coupling to the vector, we construct flows between different values of $L$ as well
as $z$.  For concreteness, we take a simple cubic potential (so the scalar can flow between
two critical points), although the general features of the flows do not depend on the details
of the potential.

This paper is organized as follows. In section~\ref{sec:constraints}, we introduce flow functions
$L(r)$ and $z(r)$ that reduce to constants $L_0$ and $z_0$ at Lifshitz fixed points.  We then
study the restrictions imposed on these functions due to the null energy condition.  In
section~\ref{sec:bhflow}, we emphasize the distinction between Lifshitz to Lifshitz flows versus
Lifshitz black holes.  In section~\ref{sec:pheno}, we study flows in the massive vector model
coupled to a real scalar and construct several numeral examples with $L$ and $z$ flowing
in various directions.  Finally, in section~\ref{sec:discussion}, we make a connection between
the present investigation of Lifshitz flows and related results for backgrounds that are only
conformal to Lifshitz (namely those exhibiting hyperscaling violation).

\section{Constraints on Lifshitz flows from the null energy condition}
\label{sec:constraints}

At a fixed point, a Lifshitz scaling background may be written in the form (\ref{metric}), where $L$
and $z$ determine the fixed point.  To study the flow of $L$ and $z$, we may consider asymptotic
Lifshitz spacetimes which arise from a generic bulk action
\bea
S=\fft{1}{2\kappa ^2} \int d^{d+1} x \sqrt{-g} (R+ \mathcal{L}_{\rm matter}),
\eea
where $\mathcal L_{\rm matter}$ will be taken to satisfy the null energy condition.  Extending
(\ref{metric}) into the interior of the bulk spacetime, we make the ansatz
\bea
ds_{d+1}^2 &=& -e^{2A(r)}dt^2+e^{2B(r)} d\vec x_{d-1} ^2 + dr^2.
\label{eq:bulkmetric}
\eea
In order to match with (\ref{metric}) in the UV, the functions $A$ and $B$ must satisfy
\begin{equation}
A\to z_{\rm UV}r/L_{\rm UV}\quad\mbox{and}\quad B\to r/L_{\rm UV}\qquad\mbox{as}\quad r\to\infty,
\label{eq:UVasymp}
\end{equation}
where $L_{\rm UV}$ and $z_{\rm UV}$ are constant values in the UV.

When considering flows between fixed points, we would like to extend $L$ and $z$ away from
their constant fixed point values.  A natural way to do this is to rewrite the asymptotic condition
(\ref{eq:UVasymp}) as
\begin{equation}
A'(r)\to z_{\rm UV}/L_{\rm UV}\quad\mbox{and}\quad B'(r) \to 1/L_{\rm UV}
\qquad\mbox{as}\quad r\to\infty.
\end{equation}
In particular, the derivatives $A'$ and $B'$ approach constants at the UV fixed point.  This
suggests that we define the $L(r)$ and $z(r)$ flow functions:
\bea
L(r) & \equiv & 1/B'(r), \nn \\
z(r) & \equiv & A'(r)/B'(r).
\label{eq:Lzdef}
\eea
It is clear that $L(r)$ and $z(r)$ agree with the constant Lifshitz radius $L$ and critical exponent
$z$ at Lifshitz fixed points of the flow.

We are now able to examine the consequences of the null energy condition.  Since we consider
a conventional Einstein-Hilbert action in the bulk, the null energy condition is equivalent to the
Ricci condition (\ref{eq:riccicond}).  By taking null vectors $k^\mu\partial_\mu=e^{-A}\partial_t
+\partial_r$ and $k^\mu\partial_\mu=e^{-A}\partial_t+e^{-B}\partial_x$, we obtain two conditions
\cite{Hoyos:2010at,Myers:2010tj}
\bea
R^r{}_r-R^t{}_t &=&(d-1)[-B''-B'^2+A'B']\geq 0,\nn \\
R^x{}_x-R^t{}_t &=&A''-B''+A'^2-(d-1)B'^2+(d-2)A'B'\geq 0.
\label{eq:lifshitzricci}
\eea
Note that these conditions may be written as
\begin{eqnarray}
\left[-e^{(B-A)}B'\right]'&\ge&0,\nonumber\\
\left[e^{A+(d-1)B}(A'-B')\right]'&\ge&0.
\label{nec}
\end{eqnarray}
As a result, the two functions
\begin{equation}
C_1(r)\equiv-e^{(B-A)}B',\qquad C_2(r)\equiv e^{A+(d-1)B}(A'-B'),
\label{eq:C1C2}
\end{equation}
are non-decreasing along Lifshitz flows to the UV.  The function $C_1$ was previously obtained
in \cite{Hoyos:2010at}, where it was related to the monotonicity of the speed of light in the bulk,
and hence to causality in Lifshitz holography.

While these two $c$-functions appear to be the natural monotonic quantities along Lifshitz flows,
unfortunately their relationship to $L$ and $z$ are not entirely obvious.  Ideally, one would use
(\ref{eq:Lzdef}) to rewrite $C_1$ and $C_2$ in terms of $L$ and $z$.  However, the exponential
factors preclude any straightforward interpretation.  Note that, in the relativistic case where
$A=B$, these functions reduce to $C_1=-B'=-1/L$ and $C_2=0$.  The $C_2$ function is trivial
because of Lorentz invariance between $x$ and $t$, while monotonicity of $C_1$ implies the well
known result that $L$ increases monotonically towards the UV.

In the Lifshitz case, the functions $C_1$ and $C_2$ do not approach constant values
at fixed points.  Instead it is easy to see that
\bea
C_1\sim-\fft{1}{L_0} e^{(1-z_0)r/L_0},\qquad
C_2\sim\fft{z_0-1}{L_0} e^{(z_0+d-1)r/L_0}.
\eea
In particular, both functions scale exponentially with $r$.  Of course, we could directly rewrite
the conditions (\ref{eq:lifshitzricci}) in terms of the effective $L$ and $z$ functions (\ref{eq:Lzdef}).
The result is
\begin{equation}
L'+(z-1)\ge0,\qquad Lz'-(z-1)L'+(z-1)(z+d-1)\ge0.
\label{eq:lzcond}
\end{equation}
Again, taking the relativistic case $(z=1)$ yields $L'\ge0$ for the first inequality.  However, in
general, the first inequality may be rewritten as
\begin{equation}
L'\ge-(z-1).
\label{eq:lprime}
\end{equation}
As a result, this no longer leads to a restriction on the sign of $L'$ whenever the effective critical
exponent is greater than one.  Furthermore, by combining the two inequalities, we see that
\begin{equation}
z'\ge-(z-1)(2z+d-2)/L\qquad(\mbox{provided }z\ge1),
\label{eq:zprime}
\end{equation}
so in addition the flow of $z$ can be in either direction as well.  (Although this inequality is
weaker than the individual inequalities in (\ref{eq:lzcond}), it is nevertheless straightforward to
verify that (\ref{eq:lzcond}) does not preclude flows of $L$ and $z$ in either direction.)
Finally, although the null energy condition does not appear restrictive for Lifshitz flows, it
does yield the standard requirement that $z\ge1$ at Lifshitz fixed points simply by setting
$L'=z'=0$ in (\ref{eq:lzcond}).

\section{Lifshitz versus black hole flows}
\label{sec:bhflow}

While the null energy condition yields the two constraints (\ref{eq:lzcond}) on Lifshitz flows,
the physical implications of these constraints is somewhat obscure.  To develop a better
understanding of the constraints, in the next section we will examine flows in a simple model
with a real scalar coupled to a massive gauge field.  However, before we do so, it is worth
making a distinction between AdS black holes and Lifshitz flows.

Since both holographic Lifshitz backgrounds and planar AdS black hole metrics can be written
in the form (\ref{eq:bulkmetric}), where the symmetry between time and space is broken,
solutions to
the bulk equations of motion by themselves cannot distinguish between the two cases.  Thus
the difference necessarily arises by imposing conditions on the solution.  For flows between
two Lifshitz fixed points (or between AdS and a Lifshitz fixed point), the bulk metric must flow
asymptotically between two regions with well-defined scaling, while for black holes, the IR
flow will reach a horizon and then a singularity.

\subsection{Schwarzschild-AdS black holes}

As an example of a black hole flow, consider for simplicity the pure Schwarzschild-AdS
black hole, conventionally written as
\begin{equation}
ds_{d+1}^2=L_0^2\left[-r^2fdt^2+r^2d\vec x_{d-1}^2+\fft{dr^2}{r^2f}\right],
\end{equation}
where $f=1-(r_0/r)^d$.  Although this metric is not in the form (\ref{metric}), a simple coordinate
transformation
\begin{equation}
r\to r_0\left(\cosh\fft{r}{2L_0/d}\right)^{2/d},
\end{equation}
brings it to the form
\begin{equation}
ds_{d+1}^2=L_0^2r_0^2\left(\cosh\fft{r}{2L_0/d}\right)^{4/d}\left(
-\tanh^2\fft{r}{2L_0/d}dt^2+d\vec x_{d-1}^2\right)+dr^2,
\end{equation}
where the horizon is at $r=0$ and the boundary is at $r=\infty$.
Using the definitions (\ref{eq:Lzdef}), we read off the effective quantities
\begin{equation}
L(r)=L_0\coth\fft{r}{2L_0/d},\qquad z(r)=1+\fft{d}2\mathrm{csch}^2\fft{r}{2L_0/d}.
\end{equation}
By construction, both $L$ and $z$ start at an AdS fixed point in the UV
\begin{equation}
(L,z)\stackrel{r\to\infty}{\longrightarrow}(L_0,1).
\end{equation}
However, their effective values both diverge as $r$ approaches the black hole horizon.
Of course, the Schwarzschild-AdS solution represents a thermal background,
and not a flow to a non-relativistic IR fixed point.  Thus this divergence is not surprising.

\begin{figure}[t]
\includegraphics{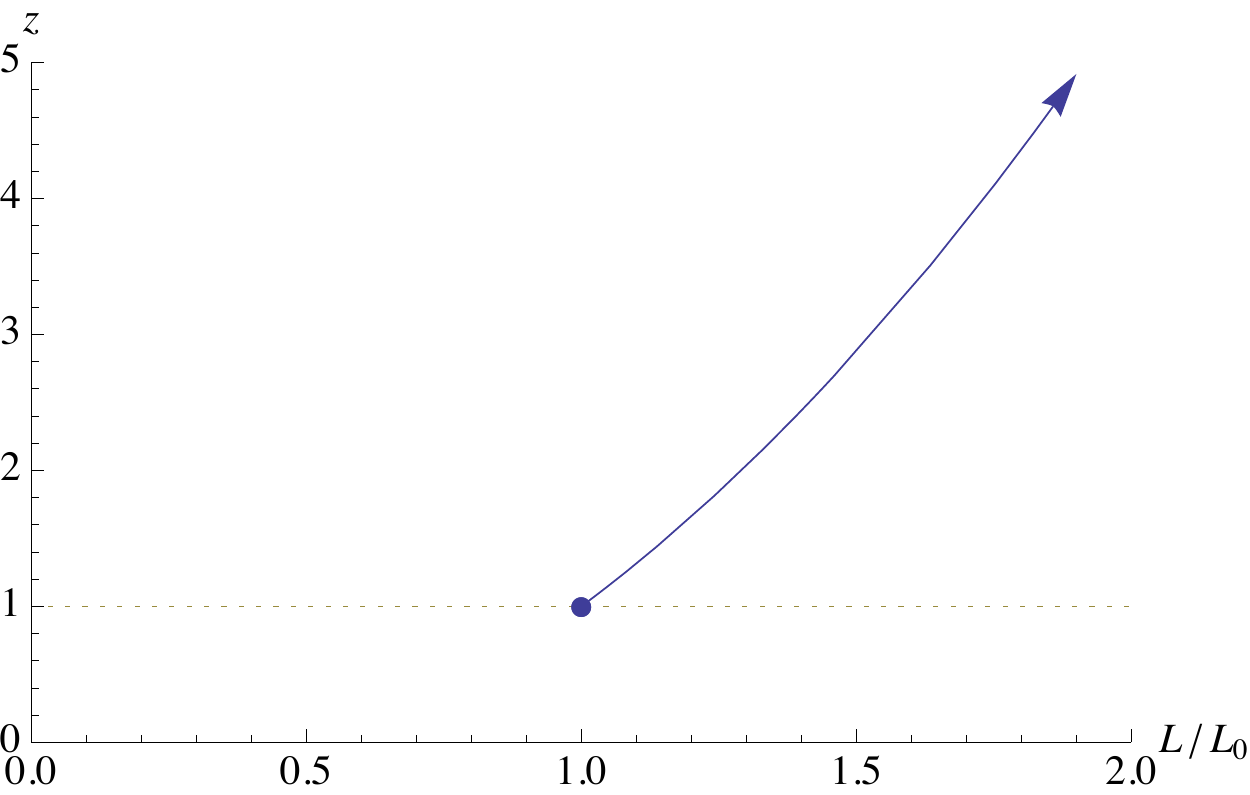}
\caption{The flow of $(L,z)$ for a pure Schwarzschild-AdS$_4$ background.  The arrow
points towards the IR ({\it i.e.}~decreasing values of $r$), and the dotted line highlights the
AdS value $z=1$.}
\label{fig:sadsbh}
\end{figure}

It is interesting to note that, while the relativistic $c$-theorem implies that $L$ is monotonically
decreasing along flows to the IR, the Schwarzschild-AdS solution instead has an increasing flow
to the IR, as shown in Fig.~\ref{fig:sadsbh}.  Since there is no matter, this flow saturates the
inequalities (\ref{eq:lprime}) and (\ref{eq:zprime}) implied by the null energy condition.  In fact,
this first inequality, $L'\ge-(z-1)$, suggests that whenever $z>1$ (which is the usual Lifshitz
case), the flow of $L$ tends to increase towards the IR.  In order for $L$ to decrease, the
bulk matter must contribute enough energy density to overcome the pull from the critical
exponent.  This balance between tendency towards non-extremality versus added matter
must be resolved in order to flow to a stable IR fixed point.  When this is done, as we will
see below, flows can then occur in directions of both increasing and decreasing $L$.

\subsection{Asymptotically Lifshitz black holes}

For actual examples of black holes in an asymptotically Lifshitz spacetime, we may consider
the model of \cite{Kachru:2008yh}, or equivalently \cite{Taylor:2008tg}.  The latter formulation
is given in terms of a massive vector field coupled to gravity
\begin{equation}
S=\fft{1}{2\kappa ^2} \int d^{d+1}x \sqrt{-g}\left( R-V -\ft14F_{\mu \nu}F^{\mu \nu}
-\ft12{m^2}A_\mu A^\mu \right).
\label{eq:mvact}
\end{equation}
Here $V$ and $m$ are constants parametrizing the model.  This system admits Lifshitz
backgrounds of the form (\ref{metric}), where $V$ and $m$ are related to $(L,z)$ by
\cite{Taylor:2008tg,Bertoldi:2009vn,Braviner:2011kz}
\begin{equation}
VL^2=-[z^2+(d-2)z+(d-1)^2],\qquad m^2L^2=(d-1)z,
\label{eq:Vmcond}
\end{equation}
and with the vector field
\begin{equation}
A=e^{zr/L}\sqrt{\fft{2(z-1)}z}dt.
\label{eq:Asoln}
\end{equation}

Black hole flows of this system were considered in \cite{Bertoldi:2009vn}, while flows
between AdS and/or Lifshitz fixed points were considered in \cite{Kachru:2008yh,Braviner:2011kz}.
The reason this simple model admits flows between Lifshitz fixed points is that the
conditions (\ref{eq:Vmcond}) are invariant under the transformation
\begin{equation}
z\to\fft{(d-1)^2}{z},\qquad L\to(d-1)\fft{L}{z}.
\label{eq:Lztrans}
\end{equation}
Hence, for an appropriate set of parameters $V$ and $m$ [corresponding to $1\le z\le(d-1)^2$],
the system admits two Lifshitz
solutions along with vacuum AdS.  As shown in \cite{Braviner:2011kz}, a relevant
deformation from the $z<d-1$ Lifshitz branch can cause a flow to either the $z>d-1$
Lifshitz branch or AdS in the IR.

\begin{figure}[t]
\includegraphics{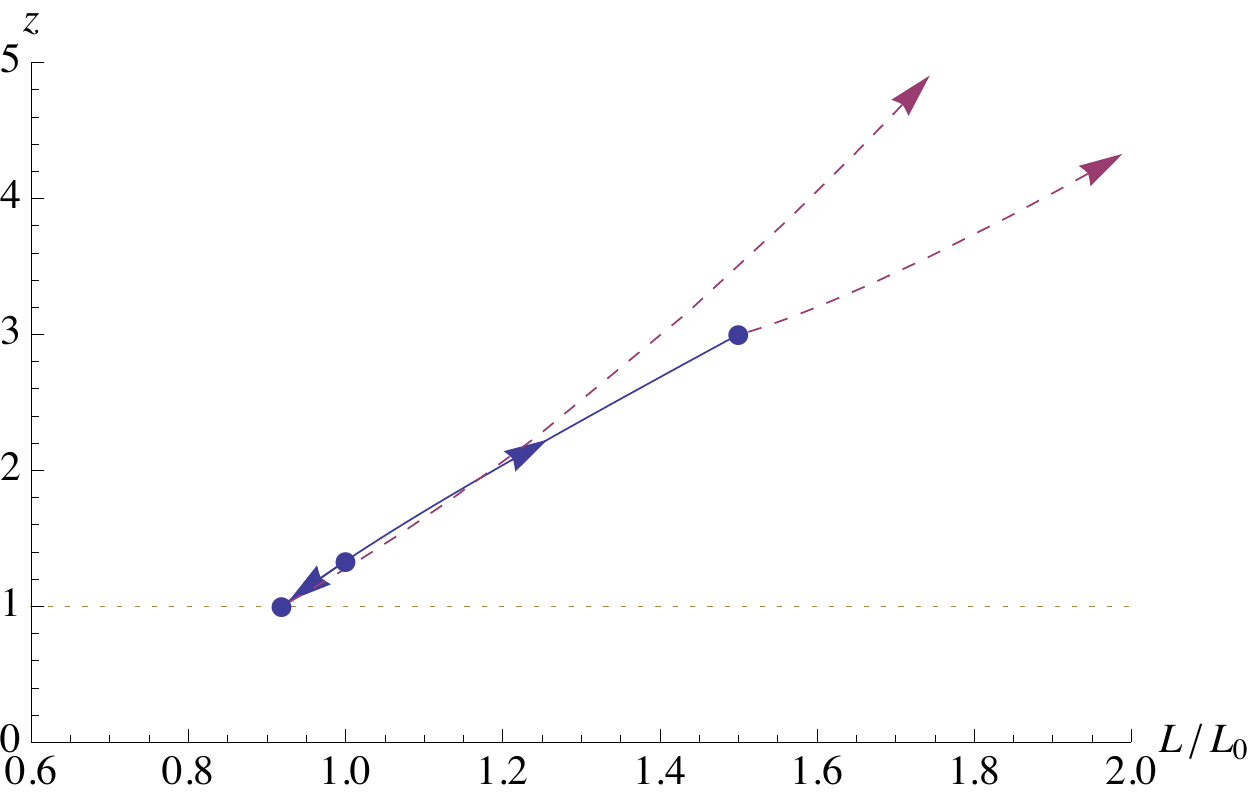}
\caption{The flow of $(L,z)$ for the massive vector model with four bulk dimensions and with
fixed points at $z=1$ (AdS), $z=4/3$ and $z=3$.  The solid line flows connect fixed points,
while the dashed lines correspond to $z=1$ (Schwarzschild-AdS) and $z=3$ black hole solutions.}
\label{fig:z=3flows}
\end{figure}

As an example of the different possible holographic flows in this model, consider a
four-dimensional bulk with $VL_0^2=-64/9$ and $m^2L_0^2=8/3$.  These quantities
are chosen so that the system admits two Lifshitz fixed points according to (\ref{eq:Vmcond})
in addition to a vacuum AdS fixed point:
\begin{equation}
(L,z)=(\sqrt{27/32}L_0,1),\qquad (L,z)=(L_0,4/3),\qquad(L,z)=(3L_0/2,3).
\end{equation}
Starting at $z=4/3$ in the UV, this background can flow to either $z=1$ or $z=3$ in the IR
\cite{Braviner:2011kz}.  The $z=1$ fixed point is of course an AdS bulk, which in turn can
be deformed into Schwarzschild-AdS.  Similarly, the $z=3$ Lifshitz background admits a
black hole \cite{Bertoldi:2009vn} with a regular horizon, and with the massive vector field
vanishing at the horizon.  These flows are shown in Fig.~\ref{fig:z=3flows}.

What this example demonstrates is that flows between fixed points can proceed in either
direction of increasing and decreasing $L$ and $z$.  For $z\approx1$, the two inequalities
(\ref{eq:lprime}) and (\ref{eq:zprime}) favor flows where both $L$ and $z$ are decreasing
towards the IR, as in the flow from Lifshitz to AdS.  However, for $z>1$, the right hand sides
of the inequalities are negative, and there is plenty of room for flows in the opposite
direction, as evidenced by the flow to $z=3$ in the IR.

In this simple example of Lifshitz backgrounds supported by a massive vector field, the
fixed point criteria (\ref{eq:Vmcond}) restricts the critical values of $(L,z)$ to lie on the curve
\begin{equation}
z=\left(\fft{m^2}{d-1}\right)L^2.
\end{equation}
As a result, flows connecting two Lifshitz fixed points in this model always proceed from
smaller $(L,z)$ in the UV to larger $(L,z)$ in the IR.  In particular, both $L$ and $z$ necessarily
flow in the same direction.  In order to examine more general Lifshitz flows, we will need
to decouple the fixed point values of $L$ and $z$.  One way to do this is to introduce a
variable effective mass for the vector field by introducing a scalar coupling.  This is what we
will do in the next section.

\section{Lifshitz flows in a phenomenological model}
\label{sec:pheno}

As we have seen above, flows between Lifshitz fixed points can have a very rich structure,
and do not appear to be strongly constrained by the null energy condition.  Flows
can also occur between AdS and Lifshitz fixed points, and between different AdS solutions, as
shown in \cite{Braviner:2011kz} in the context of six-dimensional $\mathcal N=(1,1)$
gauged supergravity and its consistent truncation.  The advantage of working in the
context of supergravity is that it naturally provides a stringy context for the bulk dual.  However,
the details of the full supergravity may at times obscure the key features of the flows.  Thus
we turn to a phenomenological model that nevertheless allows for a wide range of possibilities.

In order to examine holographic flows between unrelated fixed points, we extend the massive
vector model of \cite{Kachru:2008yh,Taylor:2008tg} by adding a real scalar $\phi$, so that the
bulk action takes the form
\begin{equation}
S=\fft1{2\kappa^2}\int d^{d+1}x\sqrt{-g}\left(R-\ft12\partial_\mu\phi\partial^\mu\phi
-\ft14F_{\mu\nu}F^{\mu\nu}-V(\phi)-W(\phi)A_\mu A^\mu\right).
\label{eq:phiAact}
\end{equation}
For the moment, we leave the potential $V(\phi)$ and scalar coupling $W(\phi)$ arbitrary,
although we assume
$\phi$ approaches a constant, $\phi(r)=\phi_0$, at fixed points of the flow.  At such points, this
system effectively reduces to (\ref{eq:mvact}) with $V=V(\phi_0)$ and $m^2=2W(\phi_0)$.
The fixed point values of $(L,z)$ can then be extracted by inverting (\ref{eq:Vmcond}).

We proceed by making a domain wall ansatz
\bea
&&ds_{d+1}^2 = -e^{2A(r)}dt^2+e^{2B(r)} d\vec x_{d-1} ^2 + dr^2, \nn \\
&&A = e^{G(r)} dt,\qquad\phi=\phi(r).
\eea
The action (\ref{eq:phiAact}) gives rise to four dynamical equations
\bea
0&=&(d-1)[B''+B'(B'-A')]+\ft12\phi'^2+e^{-2A+2G}W(\phi), \nn \\
0&=&A''+(d-2)B''+A'(A'-B')+\ft12\phi'^2-\ft12e^{-2A+2G}G'^2,\nn\\
0&=&G''+G'\left[G'-A'+(d-1)B' \right] -2W(\phi), \nn \\
0&=&\phi ''+\phi '\left[A'+(d-1)B' \right] -\partial V/\partial \phi +e^{-2A+2G}
\partial W/\partial \phi,
\eea
along with one constraint
\begin{equation}
(d-1)B'[2A'+(d-2)B']+\ft{1}{2}e^{-2A+2G}G'^2-e^{-2A+2G}W(\phi) -\ft12\phi '^2 +V(\phi) =0.
\end{equation}
The exponential factors $e^{-2A+2G}$ are associated with contractions of the vector potential,
and may be removed by defining
\begin{equation}
H(r)\equiv e^{-A(r)+G(r)}.
\end{equation}
Note that $H(r)$ is simply the time component of the vector field in the natural vielbein basis.

In order to focus on holographic RG flows, we may replace the metric functions $A$ and $B$
by the effective $L$ and $z$ functions given by (\ref{eq:Lzdef}).  With this substitution, we obtain
the equations for $(L,z,\phi,H)$
\bea
L' &=& -(z-1) +\fft{L^2}{2(d-1)} (\phi '^2 +2W(\phi) H^2),\nn\\
z' &=& -\fft{(z-1) (2z +d-2)}{L} +\fft{L}{2} \left[ \fft{z-1}{d-1}\phi'^2+\left(H'+\fft{z}{L} H \right)^2
+\fft{z+d-2}{d-1}2W(\phi)H^2 \right],\nn\\
\phi '' &=& - \fft{z+d-1}{L} \phi '+\fft{\partial V}{\partial \phi} -\fft{\partial W}{\partial \phi} H^2,
\nn\\
H'' &=& -\fft{z+d-1}{L}H'-\left[ \left( \fft{z}{L} \right) ' +\fft{(d-1)z}{L^2} \right]H
+ 2W(\phi) H.
\label{eqn_eoms}
\eea
In these variables, the constraint equation becomes
\begin{equation}
V(\phi)-W(\phi)H^2-\fft{1}{2}\phi '^2 +\fft{1}{2}\left(H'+\fft{z}{L}H\right)^2+\fft{(d-1)(2z +d-2)}{L^2} =0.
\label{eq:constraint}
\end{equation}
At this point, it is worth pointing out that the first terms on the right hand sides of the $L'$ and $z'$
equations saturate the null energy condition inequalities (\ref{eq:lprime}) and (\ref{eq:zprime}),
as they originate from the gravitational sector.  The additional terms on the right hand side
arise from the matter sector, and will contribute positively provided $z\ge1$ [as already noted
in (\ref{eq:zprime})] and $W(\phi)\ge0$.  This latter condition on $W(\phi)$ is a direct
consequence of the null energy condition.

\subsection{Lifshitz fixed points}

Before proceeding to flows, we examine fixed points of this model where $(L,z,\phi,H)$ are
all constant
\begin{equation}
(L,z,\phi,H)\to(L_0,z_0,\phi_0,H_0).
\end{equation}
Substituting these constant values into the equations of motion (\ref{eqn_eoms}) and the
constraint equation (\ref{eq:constraint}) gives two classes of solutions.  The first has vanishing
vector potential, $H_0=0$, and hence yields an AdS solution with $z_0=1$ and the AdS
radius $L_0$ related to the critical value of the potential in the usual manner
\begin{equation}
z_0=1,\qquad V(\phi_0)=-d(d-1)/L_0^2,\qquad \partial_\phi V(\phi_0)=0,\qquad H_0=0.
\label{eq:AdSfp}
\end{equation}
The second class is of Lifshitz form with
\begin{eqnarray}
V(\phi_0)L_0^2&=&-[z_0^2+(d-2)z_0+(d-1)^2],\nn\\
2W(\phi_0)L_0^2&=&(d-1)z_0,\nn\\
H_0^2&=&\fft{2(z_0-1)}{z_0},\nn\\
\partial_\phi V(\phi_0)&=&\fft{2(z_0-1)}{z_0}\partial_\phi W(\phi_0).
\label{static_constraint}
\end{eqnarray}
Note that the first two expressions map directly onto (\ref{eq:Vmcond}), while the value of
$H_0^2$ corresponds to (\ref{eq:Asoln}).  The final expression in (\ref{static_constraint})
simply ensures that the scalar is at a critical point of its effective potential (which
includes its coupling to the background vector).

Flows between fixed points may be generated by perturbing by a relevant deformation.  In order
to examine what deformations are allowed, we perform a linear stability analysis around a
given Lifshitz fixed point specified by $(L_0,z_0,\phi_0,H_0)$.  To do so, we first rewrite the
equations of motion (\ref{eqn_eoms}) in first order form by splitting the second order equations
for $\phi$ and $H$ into pairs of equations for $(\phi,\phi')$ and $(H,H')$, respectively.  We then
perturb around the Lifshitz fixed point by taking
\bea
L&=& L_0 + \epsilon \hat{L},\qquad
\phi=\phi_0+\epsilon \hat{\phi },\qquad
H=H_0 + \epsilon \hat{H}, \nn \\
z&=&z_0 + \epsilon \hat{z},\kern2.2em
\phi'=\epsilon \hat{\phi '},\kern3.7em
H'= \epsilon \hat{H'}.
\eea
For $\epsilon\ll1$, the linearized equations of motion reduce to $\mathcal{V}'=\mathcal{M}\mathcal{V}$,
where
\begin{equation}
\mathcal{V} = \lbrace \hat{L}, \hat{z}, \hat{\phi}, \hat{\phi '}, \hat{H}, \hat{H'} \rbrace ^{T},
\end{equation}
and
\begin{equation}
\mathcal M=\begin{pmatrix}
\fft{2(z_0-1)} {L_0} & -1 & \fft{L_0 ^2}{d-1}H_0^2W_1 & 0 &z_0H_0& 0 \\
\fft{2(z_0-1)(z_0+ d-2)}{L_0^2} & -\fft{z_0 +d-1}{L_0} & \fft{(z_0+d-2) L_0}{d-1}H_0^2W_1 & 0
& \fft{2z_0+d-2}{L_0}z_0H_0 &z_0H_0 \\
0 & 0 & 0 & 1 & 0 & 0 \\
0 & 0 & V_2-H_0^2W_2& -\fft{z_0 +d-1}{L_0} & -2H_0W_1 & 0 \\
0 & 0 & 0 & 0 & 0 & 1 \\
\fft{2(z_0+d-2)}{L_0^3}H_0 & 0 & \fft{z_0 +d-2}{(z_0-1) (d-1)}H_0^3W_1 & 0 &
 -\fft{2(z_0-1)(z_0 +d-2)}{L_0^2} & -\fft{3z_0 +d-3}{L_0} \end{pmatrix}.
\end{equation}
Here we have expanded the potential $V(\phi)$ and scalar coupling $W(\phi)$ about the critical
value $\phi_0$:
\begin{eqnarray}
V(\phi)&=&V_0+V_1(\phi-\phi_0)+\ft12V_2(\phi-\phi_0)^2+\cdots,\nonumber\\
W(\phi)&=&W_0+W_1(\phi-\phi_0)+\ft12W_2(\phi-\phi_0)^2+\cdots.
\label{eq:VWexpand}
\end{eqnarray}

At this point, it is worth summarizing the various parameters that enter into the matrix $\mathcal M$.
The critical point determined by (\ref{static_constraint}) may be thought of as having particular
values of the Lifshitz parameters $(L_0,z_0)$.  Then, at this fixed point, the constant values of
$V(\phi_0)$ and $W(\phi_0)$ become
\begin{equation}
V_0=-\fft{z_0^2+(d-2)z_0+(d-1)^2}{L_0^2},\qquad
W_0=\fft{(d-1)z_0}{2L_0^2}.
\label{eq:V0W0}
\end{equation}
These quantities, however, do not directly enter into $\mathcal M$.  The linear terms in
(\ref{eq:VWexpand}) are constrained by the last condition in (\ref{static_constraint})
\begin{equation}
V_1=H_0^2W_1,\qquad\mbox{where}\qquad H_0^2=\fft{2(z_0-1)}{z_0}.
\label{eq:V1W1}
\end{equation}
Since $H_0$ is determined in terms of $z_0$, it cannot considered a free parameter.  As a result,
the linearized behavior at a Lifshitz fixed point depends only on the values of $(L_0,z_0)$, the
linear potential parameter $W_1$ (or equivalently $V_1$) and the quadratic potential parameters
$V_2$ and $W_2$.

The solution to the first order equation $\mathcal V'=\mathcal M\mathcal V$ is given by
\begin{equation}
\mathcal V(r)=\sum_i\mathcal V_ie^{\lambda_ir},
\end{equation}
where $\{\lambda_i\}$ are the eigenvalues of $\mathcal M$, with corresponding eigenvectors
$\{\mathcal V_i\}$.  Since we assume the asymptotic form of the metric (\ref{metric}), where
$r\to\infty$ corresponds to the UV, relevant deformations (which induce a flow to the IR) will
correspond to negative eigenvalues, namely $\lambda^{\rm (UV)}_i<0$.  If this flow terminates at a stable
IR fixed point, then it will necessarily approach the IR fixed point along a direction or set of
directions with positive eigenvalues, $\lambda^{\rm (IR)}_i>0$.

The eigenvalues of $\mathcal M$ are easily determined, although their general expressions are
not particularly illuminating.  Much of the complication arises when the linear term is non-vanishing
({\it i.e.}~when $W_1\ne0$), as in this case the scalar and vector fluctuations no longer decouple at
fixed points.  We therefore restrict the presentation of explicit results to the case when $W_1=0$.  In
this case, there is a single marginal mode, $\lambda_1=0$, with eigenvector
\begin{equation}
\mathcal V_1=\left(H_0L_0,2z_0H_0,0,0,\fft2{z_0},0\right).
\end{equation}
This marginal direction corresponds to holding $\phi$ fixed, and moving along the curve given by
$2W_0L_0^2=(d-1)z_0$ in (\ref{static_constraint}).  Note, however, that while this deformation is
marginal in the set of first order equations, movement along this direction will not satisfy the
constraint (\ref{eq:constraint}).  Thus this mode takes us outside of a given model, and in particular
shifts the vacuum energy $V_0$ (which is a constant of integration of the set of first order equations).

There are three additional modes that keep $\phi$ fixed, but involve a combination of the metric and
vector field.  The first one of this set is always relevant, with $\lambda_2=-(z_0+d-1)/L_0$ and
eigenvector
\begin{equation}
\mathcal V_2=\left(\fft{H_0L_0}{z_0+d-2},\fft{z_0-d+1}{z_0+d-2}H_0,0,0,-\fft{2}{z_0},
\fft{2(z_0+d-1)}{z_0L_0}\right).
\end{equation}
The other two are paired up, with
\begin{equation}
\lambda_{3,4}=-\fft{z_0+d-1}{2L_0}(1\pm\Delta),
\label{eq:lambda34}
\end{equation}
and
\begin{eqnarray}
\mathcal V_{3,4}&=&\biggl(\fft{2-4z_0+(z_0+d-1)(1\mp\Delta)}{2(z_0+d-2)}H_0L_0,
\fft{-4z_0+(z_0+d-1)(1\mp\Delta)}2H_0,\nonumber\\
&&\qquad0,0,-\fft2{z_0},\fft{(z_0+d-1)(1\mp\Delta)}{z_0L_0}\biggr),
\end{eqnarray}
where
\begin{equation}
\Delta=\sqrt{1+\fft{8(z_0-1)(z_0-d+1)}{(z_0+d-1)^2}}.
\end{equation}
Since we take $\Delta$ to be positive, $\lambda_3$ [corresponding to the top sign in
(\ref{eq:lambda34})] will always be negative, corresponding to a relevant deformation.
However, the behavior of $\lambda_4$ depends on the value of $\Delta$.  For $1\le z_0\le d-1$,
we find $\Delta\le1$, so that $\lambda_4\le0$.  On the other hand, we obtain
an irrelevant deformation, $\lambda_4>0$, whenever $z_0>d-1$.

The remaining two modes only involve the scalar field $\phi$ and its derivative (in the linearized
analysis), and have eigenvalues
\begin{equation}
\lambda_{5,6}=-\fft{z_0+d-1}{2L_0}(1\pm\Sigma),
\end{equation}
along with eigenvectors
\begin{equation}
\mathcal V_{5,6}=\left(0,0,\fft{(z_0+d-1)(1\mp\Sigma)}{2L_0},V_2-H_0^2W_2,0,0\right),
\end{equation}
where
\begin{equation}
\Sigma=\sqrt{1+\fft{4L_0^2(V_2-H_0^2W_2)}{(z_0+d-1)^2}}.
\end{equation}
Since we are assuming the absence of linear terms in $V(\phi)$ and $W(\phi)$ at the critical point,
it is clear that the combination $m_{0,\rm eff}^2\equiv V_2-H_0^2W_2$ is simply the effective mass
of $\phi$.  Note that, in the AdS case when $z_0=1$, we find the expected relation between scalar
mass and conformal dimension
\begin{equation}
\lambda_{5,6}=-\fft{1}{L_0}\left(d/2\pm\sqrt{(d/2)^2+m_{0,\rm eff}^2L_0^2}\right)\qquad(\mbox{for }z_0=1).
\label{eq:E0phi}
\end{equation}
The only modification for general $z_0$ is then the replacement $d\to z_0+d-1$.
The behavior of $\lambda_{5,6}$ depends on the sign of $m_{0,\rm eff}^2$.  For $-[(z_0+d-1)/2L_0]^2
\le m_{0,\rm eff}^2\le0$ (where the lower bound is taken as a generalization of the
Breitenlohner-Freedman bound \cite{Breitenlohner:1982bm,Breitenlohner:1982jf}),
both $\lambda_5$ and $\lambda_6$ are negative, so the scalar deformations are both
relevant.  However, for $m_{0,\rm eff}^2>0$, one of the deformations becomes irrelevant,
corresponding to $\lambda_6>0$.

In summary, after linearizing about a Lifshitz fixed point, we find a set of six perturbations, of which
one is marginal.  (This remains the case even when $W_1\ne0$.)  However, this marginal
direction is to be disregarded, as it is eliminated by the constraint (\ref{eq:constraint}).  Of the
remaining five modes, three are always relevant, while the other two will depend on the
parameters of the fixed point.  In the absence of a linear coupling, the modes become irrelevant in
the following cases:
\bea
\lambda_4 >0, && \text{when} \quad z_0 > d-1, \label{H_push} \\
\lambda_6 >0, && \text{when} \quad m_{0,\rm eff}^2\equiv V_2-H_0^2W_2 >0. \label{phi_push}
\eea
Since $\mathcal V_{2,3,4}$ have vanishing components along $\hat{\phi}$ and $\hat{\phi}'$, we label
the flows induced by these modes as ``vector field driven''%
\footnote{These modes are actually combinations of vector and metric perturbations.}.
On the other hand, since
$\mathcal V_{5,6}$ initiates flows along $\hat\phi$ and $\hat\phi'$, we will label such flows as
``scalar field driven''.

\subsection{AdS fixed points}

Although our main interest is to examine Lifshitz fixed points, the simple model (\ref{eq:phiAact})
also admits AdS fixed points given by (\ref{eq:AdSfp}).  The stability analysis of such fixed points
is similar to that of Lifshitz fixed points.  However, we cannot simply set $z_0=1$ in the previous
results, as the second condition in (\ref{static_constraint}) is no longer applicable when the massive
vector is turned off.  Nevertheless, the deformations can be classified in essentially the same
manner.

Again, there is a single marginal mode with $\lambda_1=0$ and
\begin{equation}
\mathcal V_1=(1,0,0,0,0,0).
\end{equation}
This marginal direction corresponds to shifting the cosmological constant, and is again removed
by the constraint (\ref{eq:constraint}).  The three modes found above involving the metric and
vector field now split into a relevant metric deformation with $\lambda_2=-d/L_0$ and
\begin{equation}
\mathcal V_2=(L_0,d,0,0,0,0),
\end{equation}
and a pair of vector field deformations with
\begin{equation}
\lambda_{3,4}=-\fft{d\pm\sqrt{(d-2)^2+8W_0L_0^2}}{2L_0}
\end{equation}
and
\begin{equation}
\mathcal V_{3,4}=\left(0,0,0,0,-1,\fft{d\pm\sqrt{(d-2)^2+8W_0L_0^2}}{2L_0}\right).
\end{equation}
The metric deformation generates a flow to a Schwarzschild-AdS black hole, while the two
vector field modes have the expected conformal dimensions for a massive vector in AdS
with an effective mass of $m_{1,\rm eff}^2=2W_0$.  (Recall that the null energy condition
demands $W(\phi)\ge0$, so that $m_{1,\rm eff}^2$ must be non-negative.)
For $0\le m_{1,\rm eff}^2\le (d-1)/L_0^2$,
both vector modes are relevant, while for $m_{1,\rm eff}^2>(d-1)/L_0^2$, the deformation
corresponding to $\lambda_4$ becomes irrelevant.

Finally, the remaining two modes are those of the scalar field, with
\begin{equation}
\lambda_{5,6}=-\fft1{L_0}\left(d/2\pm\sqrt{(d/2)^2+V_2L_0^2}\right),
\end{equation}
and
\begin{equation}
\mathcal V_{5,6}=\left(0,0,\fft{d/2\mp\sqrt{(d/2)^2+V_2L_0^2}}{L_0},V_2,0,0\right),
\end{equation}
in agreement with (\ref{eq:E0phi}), where $m_{0,\rm eff}^2=V_2$ is just the scalar
mass read off from the potential.  As a result, we find that the deformations away from an AdS
fixed point are all relevant, except for the two cases
\begin{eqnarray}
\lambda_4 >0, && \text{when} \quad m_{1,\rm eff}^2\equiv2W_0 > (d-1)/L_0^2, \label{eq:AdSH_push} \\
\lambda_6 >0, && \text{when} \quad m_{0,\rm eff}^2\equiv V_2 >0,
\label{eq:AdSphi_push}
\end{eqnarray}
where the corresponding deformations become irrelevant.  As above, flows to or from AdS
fixed points may be thought of as either vector field driven or scalar field driven.  Of course, since
the vector field vanishes at AdS fixed points, a scalar field deformation may initiate a flow
between AdS fixed points, but would not otherwise generate a non-vanishing vector field
background.  Such scalar field driven flows will thus maintain $z=1$ throughout the flow,
unless accompanied by a metric or vector perturbation.

\subsection{Flows between fixed points}

With the linear stability analysis out of the way, we now consider flows between fixed points.
Such flows may be generated by starting at a UV fixed point $(L_{\rm UV},z_{\rm UV})$, and turning
on a relevant deformation.
This model admits five deformations away from any fixed point, of which three are always relevant.
Hence it is always possible to construct a flow away from the UV.  However, not all such flows will
terminate in an IR fixed point.  We find that generic deformations will flow to a singularity with both the
scalar and vector fields running away to infinity.  Furthermore, as discussed previously, flows to
a black hole horizon are also possible.

Assuming, however, that the flow reaches an IR fixed point with $(L_{\rm IR},z_{\rm IR})$, it will
approach this fixed point in an irrelevant direction.  For a Lifshitz fixed point (assuming vanishing
linear $\phi$ coupling at this point), the criteria for an irrelevant deformation are given by either
(\ref{H_push}), which requires $z_{\rm IR}>d-1$, or (\ref{phi_push}), which requires
$m_{0,\rm IR}^2>0$.  The former corresponds to a vector field driven flow, while the latter corresponds
to a scalar field driven flow.  For an AdS fixed point in the IR, the first criterion is replaced by
(\ref{eq:AdSH_push}), which requires $m_{1,\rm IR}^2>(d-1)/L_{\rm IR}^2$.

At this point, it is worth noting that the flows of \cite{Kachru:2008yh,Braviner:2011kz} in the
massive vector model are naturally vector field driven, as this model lacks the extra scalar.
In this case, Lifshitz fixed points in the IR must lie in the $z_{\rm IR}>d-1$ branch, as can
be seen in the example of Fig.~\ref{fig:z=3flows}.  Similarly, it can be seen that AdS fixed points
in the IR can be reached for flows starting with $1<z_{\rm UV}<(d-1)^2$.

In the absence of a specific model, we forego a general analysis of flows between fixed points.
However, it is instructive to provide a few examples of scalar field driven Lifshitz to Lifshitz flows.
In particular, we may engineer such flows by choosing appropriate forms of the scalar potential
$V(\phi)$ and vector field coupling $W(\phi)$.  A natural, and fairly minimal choice would be to
construct the potential $V$ to have two critical points, corresponding to the UV and IR.  We then
design $W$ so that the effective vector mass, $m_{1,\rm eff}^2$, takes on appropriate values
at the critical points, so as to yield the desired Lifshitz scaling according to (\ref{eq:V0W0}).
Although it is possible to choose a linear function for $W$, for simplicity we demand that both
$V$ and $W$ have critical points at the same values of $\phi$.  This allows us to avoid any
linear shift in the potential implied by (\ref{eq:V1W1}).

The simplest model with two critical points involves taking cubic potentials
\begin{eqnarray}
V(\phi)&=&V_0+V_1\phi+V_2\phi^2+V_3\phi^3,\nonumber\\
W(\phi)&=&W_0+W_1\phi+W_2\phi^2+W_3\phi^3.
\end{eqnarray}
Although a cubic $W(\phi)$ will always become negative in some domain, we will always
restrict $\phi$ so that $W(\phi)\ge0$ to ensure compatibility with the null energy condition.
This is a legitimate restriction since we are only interested in classical solutions to the equations
of motion.
The eight constants of the potentials are fixed by demanding that they admit an IR critical point at
$\phi=0$ with $(L_{\rm IR},z_{\rm IR})$ and a UV critical point at $\phi=\phi_0$ with
$(L_{\rm UV},z_{\rm UV})$.  We find, in particular
\begin{eqnarray}
V_0&=&-\fft{z_{\rm IR}^2+(d-2)z_{\rm IR}+(d-1)^2}{L_{\rm IR}^2},\nonumber\\
V_1&=&0,\nonumber\\
V_2\phi_0^2&=&3\left[\fft{z_{\rm IR}^2+(d-2)z_{\rm IR}+(d-1)^2}{L_{\rm IR}^2}
-\fft{z_{\rm UV}^2+(d-2)z_{\rm UV}+(d-1)^2}{L_{\rm UV}^2}\right],\nonumber\\
V_3\phi_0^3&=&-2\left[\fft{z_{\rm IR}^2+(d-2)z_{\rm IR}+(d-1)^2}{L_{\rm IR}^2}
-\fft{z_{\rm UV}^2+(d-2)z_{\rm UV}+(d-1)^2}{L_{\rm UV}^2}\right],\nonumber\\
\end{eqnarray}
and
\begin{eqnarray}
W_0&=&\fft{(d-1)z_{\rm IR}}{2L_{\rm IR}^2},\nonumber\\
W_1&=&0,\nonumber\\
W_2\phi_0^2&=&-\fft{3(d-1)}2\left[\fft{z_{\rm IR}}{L_{\rm IR}^2}-\fft{z_{\rm UV}}{L_{\rm UV}^2}
\right],\nonumber\\
W_3\phi_0^3&=&(d-1)\left[\fft{z_{\rm IR}}{L_{\rm IR}^2}-\fft{z_{\rm UV}}{L_{\rm UV}^2}
\right].
\end{eqnarray}
It is worth noting, however, that this model does not necessarily admit a
stable flow from arbitrary values of $(L_{\rm UV},z_{\rm UV})$ to $(L_{\rm IR},z_{\rm IR})$,
as relevant deformations from the UV could often flow to a black hole or singular geometry.
In addition, there is a two-fold degeneracy of fixed points (both at the UV and the IR),
corresponding to the map given by (\ref{eq:Lztrans}).  Hence flows could originate or terminate
in the dual fixed points from the ones that were originally desired.

\begin{figure}[t]
\includegraphics{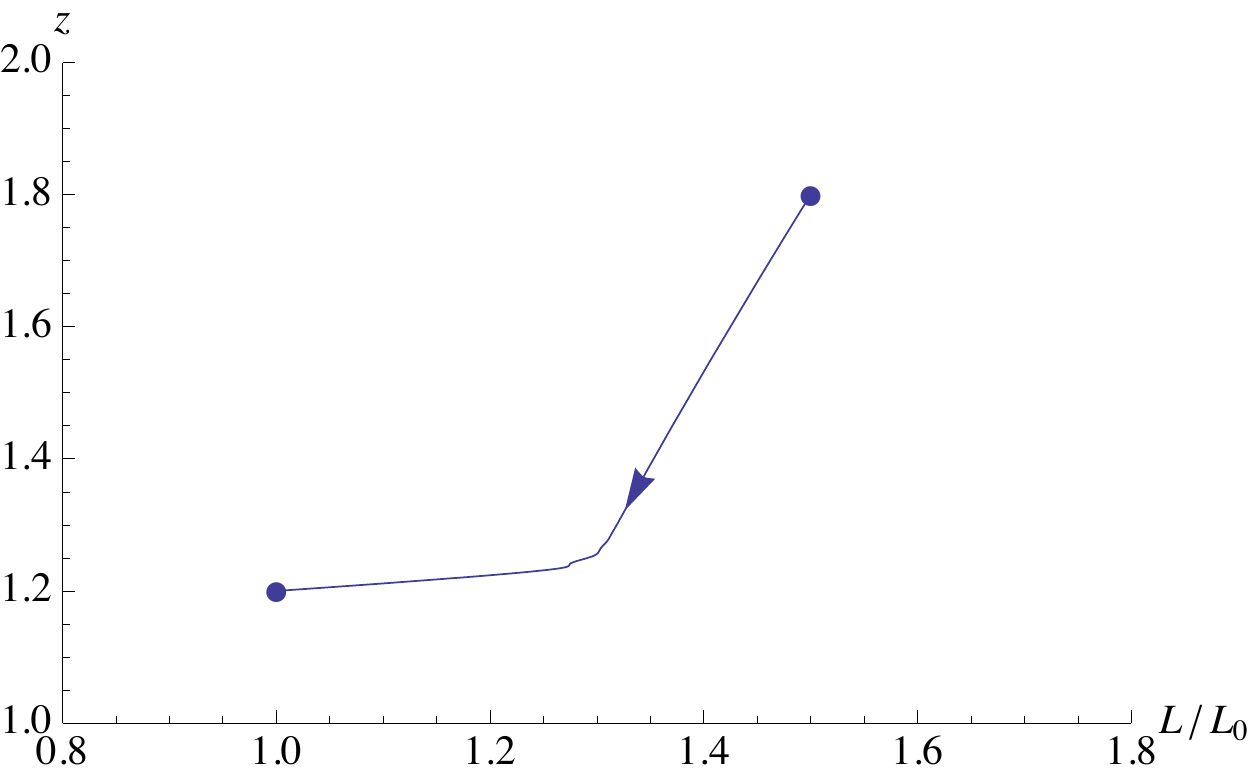}
\caption{A scalar field driven flow with decreasing $z$ and $L$.  We have taken $d=3$ and
$\phi_0=1$, along with the fixed point parameters given in (\ref{eq:fig3params}).}
\label{fignum:1}
\end{figure}

In order to go beyond the linearized analysis, we solve the equations of motion (\ref{eqn_eoms})
numerically, integrating out from the IR to the UV along an irrelevant direction.  The reason we
integrate out from the IR is that, as indicated above, there are at most only two irrelevant
deformations in the IR, given by the conditions (\ref{H_push}) and (\ref{phi_push}).  Since
we are interested in scalar field driven flows (where $\phi$ flows between the UV and IR
critical points of the potential), we naturally choose the $\lambda_6$ direction for the deformation.
An example of a scalar field driven flow towards decreasing $z$ is shown in Fig.~\ref{fignum:1},
where we have chosen a four-dimensional bulk and fixed point parameters
\begin{equation}
(L_{\rm UV},z_{\rm UV}) = (3L_0/2,9/5),\qquad (L_{\rm IR},z_{\rm IR})=(L_0,6/5).
\label{eq:fig3params}
\end{equation}
Note that $z$ lies in the range $1\le z\le d-1$ for this class of flows.  We see that there is
some slight ringing as $\phi$ flows away from $\phi_0$ (where $V(\phi_0)$ is a local maximum)
in the UV.

\begin{figure}[t]
\includegraphics{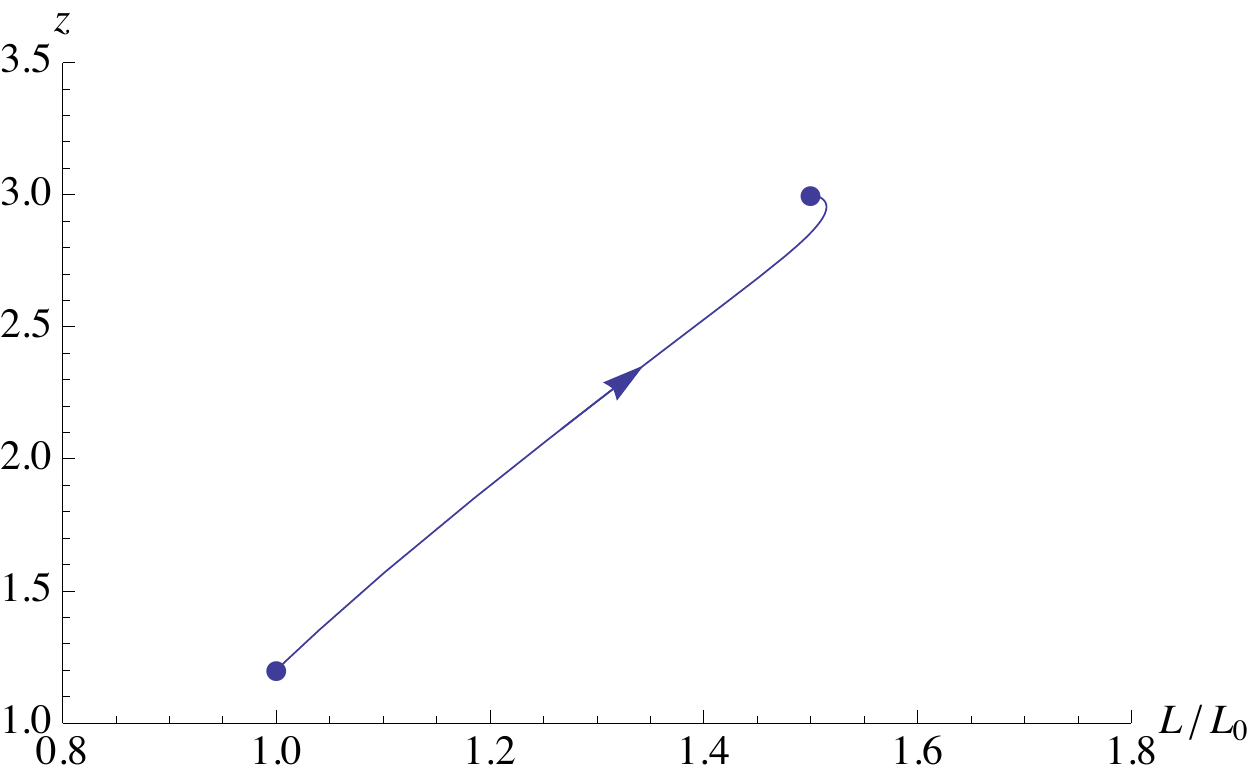}
\caption{A scalar field driven flow with increasing $z$ and $L$.  The fixed point parameters
are given in (\ref{eq:fig4params}), along with $d=3$ and $\phi_0=1$.}
\label{fignum:2}
\end{figure}

\begin{figure}[t]
\includegraphics{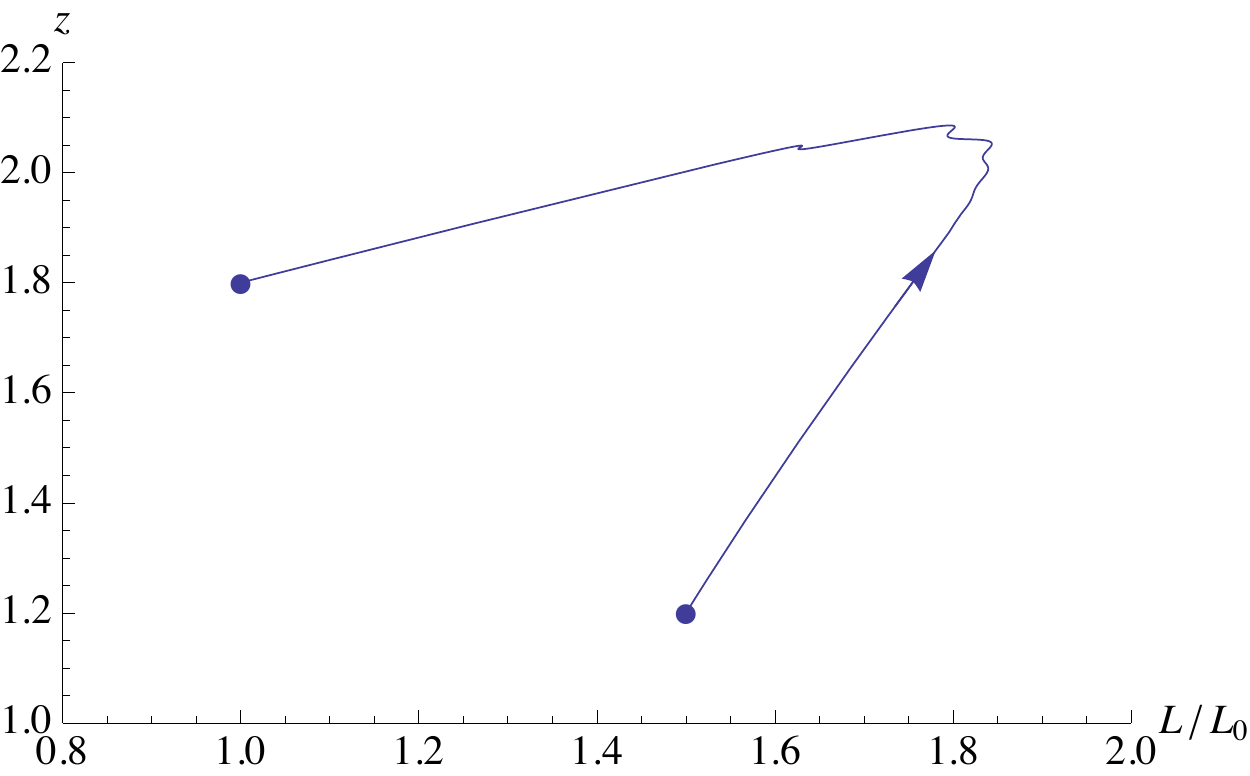}
\caption{A scalar field driven flow with increasing $z$ and decreasing $L$.  We have taken
$d=3$ and $\phi_0=1$, along with the parameters given in (\ref{eq:fig5params}).}
\label{fignum:5}
\end{figure}

It is also easy to obtain flows towards increasing $z$.  One such example is shown in
Fig.~\ref{fignum:2}, where the fixed point values are given by
\begin{equation}
(L_{\rm UV},z_{\rm UV}) = (L_0,6/5),\qquad (L_{\rm IR},z_{\rm IR})=(3L_0/2,3).
\label{eq:fig4params}
\end{equation}
The general feature of this flow is similar to the vector field driven flow from $z=4/3$ to
$z=3$ shown in Fig.~\ref{fig:z=3flows}.  However, we note that the effective value of
$L$ actually overshoots its IR fixed point value before finally turning around
and reaching it at the end of the flow.  This is a clear indication that the function $L(r)$ is
not constrained to be monotonic by the null energy condition.  It is not obvious whether this
has any physical significance, as the effective $L$ and $z$ functions are not necessarily
observable along the flow.  It is also possible to obtain flows towards increasing $z$,
but with decreasing $L$.  An interesting example is given in Fig.~\ref{fignum:5}, corresponding
to
\begin{equation}
(L_{\rm UV},z_{\rm UV}) = (3L_0/2,6/5),\qquad (L_{\rm IR},z_{\rm IR})=(L_0,9/5).
\label{eq:fig5params}
\end{equation}
Here we see that the effective values of both $L$ and $z$ flow away from the UV for some
distance before turning around and reaching the IR fixed point.  The ringing near the UV
fixed point of the potential is also more pronounced, as both the scalar and vector fields
participate in the flow.

\section{Discussion}
\label{sec:discussion}

In the relativistic case ($z=1$), the consequence of the null energy condition directly leads to
the holographic $c$-theorem, $L'\ge0$.  However, once relativistic invariance is no longer
required, the implications of the null energy condition are very much relaxed.  Although it
is possible to define two monotonic functions, $C_1$ and $C_2$, as in (\ref{eq:C1C2}),
neither one of them serves the purpose of a useful $c$-function, as they do not approach
constant values at fixed points (except when $z=1$, in which case $C_1$ coincides with
the usual holographic $c$-function).

For a Lifshitz fixed point, the more useful quantities $(L,z)$ do not exhibit any obvious
monotonicity.  Using a cubic potential model, we have constructed flows between two Lifshitz
fixed points that satisfy the null energy condition, and where $L$ and $z$ are simultaneously
increasing or decreasing or where $L$ is decreasing while $z$ is increasing.  So far, however,
we have been unable to find any examples of flows with simultaneously decreasing $z$ and
increasing $L$.  It is possible that this is a feature of the toy model, as the null energy
condition in itself does not preclude such flows.  Nevertheless, the inequalities (\ref{eq:C1C2})
obviously place some restraints on the allowed flows, and it would be interesting to see whether
there is a deeper reason why we have not found any flows where $z$ decreases while $L$
increases.

Although we have focused on flows between scale invariant Lifshitz fixed points, there has
been recent interest in systems involving hyperscaling violation (parametrized by
$\theta$) in addition to the dynamical critical exponent $z$.  Gravitational duals to such models
have been realized in Einstein-Maxwell-dilaton theories
\cite{Gubser:2009qt,Charmousis:2010zz,Perlmutter:2010qu}, and aspects of the null energy
condition have been investigated in \cite{Ogawa:2011bz,Huijse:2011ef,Dong:2012se}.
Metrics exhibiting hyperscaling violation may be written in the conformally Lifshitz form%
\footnote{We have continued to use `relativistic AdS/CFT' notation, where the bulk
is $(d+1)$-dimensional.  Most of the recent hyperscaling violation literature uses $d$ to denote
the spatial dimensions of the field theory, so that the bulk is $(d+2)$-dimensional.}
\begin{equation}
ds_{d+1}^2= r^{2\theta/(d-1)}\left[-\fft{dt^2}{r^{2z}}+\fft{d\vec x_{d-1}^{\,2}+dr^2}{r^2}\right].
\end{equation}
In order to make contact with (\ref{eq:bulkmetric}), we transform $r\to r^{(d-1)/\theta}$ (and
rescale the coordinates) so that
\begin{equation}
ds_{d+1}^2=-r^{2(1-z(d-1)/\theta)}dt^2+r^{2(1-(d-1)/\theta)}d\vec x_{d-1}^{\,2}+dr^2.
\label{eq:hypermet}
\end{equation}
Therefore, in a hyperscaling region, we make the identification
\begin{equation}
A(r)=\left(1-\fft{z(d-1)}\theta\right)\log r,\qquad
B(r)=\left(1-\fft{(d-1)}\theta\right)\log r.
\label{eq:ABhyper}
\end{equation}
This may be contrasted with the behavior $A\sim zr/L$ and $B\sim r/L$ in the case where
$\theta=0$.  (Note, however, that this identification is not well-behaved in the limit $\theta\to0$,
as the scale-invariant metric functions behave as exponentials and not power-laws.)

It is interesting to see what the flow functions $L(r)$ and $z(r)$ defined in (\ref{eq:Lzdef})
look like for the metric (\ref{eq:hypermet}).  We find
\begin{equation}
L(r)=-\fft\theta{d-1-\theta}r,\qquad z(r)=1+\fft{(z-1)(d-1)}{d-1-\theta},
\label{eq:Lzhyper}
\end{equation}
where we should keep in mind that $z(r)$ is an effective function that was designed to
match the critical exponent only at pure Lifshitz fixed points, while $z$ is the true critical
exponent given in (\ref{eq:hypermet}).  By construction, $z(r)$ and $z$ coincide in the absence
of hyperscaling violation ({\it i.e.}~when $\theta=0$).  However, the effective function $L(r)$
runs linearly in $r$, and hence cannot be assigned a fixed value in a hyperscaling region of
any flow.

For $L(r)$ and $z(r)$ given in (\ref{eq:Lzhyper}), the null energy condition (\ref{eq:lzcond})
gives rise to the inequalities
\begin{equation}
(d-1-\theta)((z-1)(d-1)-\theta)\ge0,\qquad (z-1)(z+d-1-\theta)\ge0,
\label{eq:hypernec}
\end{equation}
as noted in \cite{Dong:2012se}%
\footnote{Note that we must take $d\to d+1$ to match the notation of \cite{Dong:2012se}.}.
It would be interesting to see whether these conditions can be extended along a complete
flow between regions with different scaling behavior.  Of course, the two functions
$C_1(r)$ and $C_2(r)$ in (\ref{eq:C1C2}) are valid $c$-functions regardless of the geometry.
However, the usefulness of the functions depend on being able to express them in terms of
`physical' quantities such as $\theta$ and $z$ at fixed points of the flow.  We have, unfortunately,
not been able to find a suitable set of $\theta(r)$ and $z(r)$ functions that extend the behavior
of (\ref{eq:ABhyper}) beyond the fixed points.  The main obstacle in doing so is to properly
reproduce the power-law behavior implicit in the $\log r$ running of $A(r)$ and $B(r)$.

A possible attempt at defining a $c$-function for flows with hyperscaling violation is to capture
the power-law behavior $e^A\sim r^\alpha$ by removing the explicit $r$ dependence
in (\ref{eq:ABhyper}).  This may be done by forming ratios, such as $(A')^2/A''$, $(B')^2/B''$ or
$A'/B'$.  For example, we may define
\begin{equation}
z(r)\equiv\fft{B''+A'B'}{B''+(B')^2},\qquad \theta(r)\equiv(d-1)\fft{B''}{B''+(B')^2},
\label{eq:zthetadef}
\end{equation}
so that $z(r)$ and $\theta(r)$ approach the constant values $z_0$ and $\theta_0$ at
fixed points of the flow.  This extension of $z$ and $\theta$ away from fixed points is not
unique, but has the advantage that the first inequality in (\ref{nec}) yields
simply
\begin{equation}
(d-1-\theta(r))((z(r)-1)(d-1)-\theta(r))\ge0,
\label{eq:firstineq}
\end{equation}
which is identical in form to the first equation in (\ref{eq:hypernec}), but now must hold everywhere
along the flow.  Unfortunately, the second inequality in (\ref{nec}) does not have a simple expression
in terms of $z(r)$ and $\theta(r)$ because $A''$ in (\ref{nec}) has no natural counterpart in the
definition (\ref{eq:zthetadef}).  Note, also, that the inequality (\ref{eq:firstineq}) does not take the
conventional form of a gradient of a $c$-function, and hence does not suggest any direction for
the flow of $z$ or $\theta$.  Nevertheless, this discussion suggests that additional information
may be captured from the null energy condition beyond just the fixed point inequalities
(\ref{eq:hypernec}).

Finally, once again ignoring hyperscaling violation, we note that it may be possible to resolve
the tidal singularity at the Lifshitz horizon by flowing into AdS$_2\times\mathbb R^{d-1}$
in the deep IR \cite{Harrison:2012vy}.  Since the AdS$_2\times\mathbb R^{d-1}$ geometry
is obtained by taking $B(r)=0$ in (\ref{eq:bulkmetric}), the definition of the flow functions
$L(r)$ and $z(r)$ in (\ref{eq:Lzdef}) break down in this case.  In particular, the flow to
AdS$_2\times\mathbb R^{d-1}$ is represented by
\begin{equation}
L(r)\to\infty,\qquad z(r)\to\infty,\qquad\mbox{with $z(r)/L(r)=L_2$ held fixed},
\end{equation}
where $L_2$ is the radius of the emergent AdS$_2$.  Note that this is distinct from the
Schwarzschild black hole flows shown in Figs.~\ref{fig:sadsbh} and \ref{fig:z=3flows}, as
the Schwarzschild flows have $z(r)\sim L^2(r)\to\infty$ as the flow approaches the black hole
horizon.

Regardless of the nature of the model, as long as it is a two-derivative theory of gravity, the
power of the null energy condition is that it directly translates to a condition on the geometry,
(\ref{eq:riccicond}), and hence a condition on its geodesics and causal structure.  Although
there has been
progress in understanding the implications of the null energy condition for higher derivative
gravity, once additional terms enter the left-hand side of the Einstein equation, the null
energy condition by itself no longer completely determines the Ricci tensor.  As a result,
additional conditions may be required in the gravitational sector in order to have a well-behaved
holographic dual.  It would be worthwhile to extend our results for holographic
Lifshitz flows to the case of higher derivative gravity in the bulk, and to investigate what
form these additional conditions may take.

\acknowledgments
We wish to thank B. Burrington, C. Keeler and M. Paulos for useful discussions.  This work
was supported in part by the US Department of Energy under grant DE-FG02-95ER40899.


\end{document}